\documentclass[prb,aps,twocolumn,showkeys]{revtex4-1}

\usepackage{amssymb}
\usepackage{amsmath}
\usepackage{graphicx}
\usepackage[sort&compress]{natbib}
\usepackage[dvips]{color}
\usepackage[version=3]{mhchem}

\begin{document}

\title{Coarse-Grained Models of Biological Membranes within \\
the Single Chain Mean Field Theory}

\author{ Sergey Pogodin$^1$ and Vladimir A.
Baulin$^{1,2}$}

\affiliation{$^1$Universitat Rovira i Virgili 26 Av. dels Paisos
Catalans, 43007 Tarragona Spain}

\affiliation{$^2$ICREA, 23 Passeig Lluis Companys, 08010
Barcelona, Spain}

\keywords{\textbf{Accepted for publication in Soft Matter}
DOI:10.1039/b927437e }

\begin{abstract}
The Single Chain Mean Field theory is used to simulate the equilibrium
structure of phospholipid membranes at the molecular level. Three levels of
coarse-graining of DMPC phospholipid surfactants are present: the detailed
44-beads double tails model, the 10-beads double tails model and the minimal
3-beads model. We show that all three models are able to reproduce the
essential equilibrium properties of the phospholipid bilayer, while the
simplest 3-beads model is the fastest model which can describe adequately
the thickness of the layer, the area per lipid and the rigidity of the
membrane. The accuracy of the method in description of equilibrium
structures of membranes compete with Monte Carlo simulations while the speed
of computation and the mean field nature of the approach allows for
straightforward applications to systems with great complexity.
\end{abstract}

\maketitle

\section{Introduction}

Phospholipid membranes and self-assembled bilayers of block copolymers have
many common features originating from the amphiphilic nature of the
molecules. Phospholipids comprise of hydrophilic head group and two
hydrophobic acyl chains and the driving force of self-assembly in membranes,
as in case of block copolymers, is the hydrophobic effect \cite{Sackmann}.
Thus, theoretical methods developed for the self-assembly of block
copolymers can be applied for the description of phospholipid membranes \cite%
{Muller2}.

Theoretical description and computer simulation of phospholipid bilayers and
biological membranes is a challenging task for many decades \cite%
{Muller2,Smit}. Although phospholipids are relatively short chains in
comparison to polymers and block copolymers, the configurational space of
conformations and the number of interactions are extremely large. In
addition, the simulation of phospholipid bilayers and membranes is a problem
involving many interacting molecules, thus full description of collective
behavior would require simultaneous simulation of a large number of
molecules. That is why, full atomistic molecular dynamics models \cite%
{Tieleman,Damodaran,Damodaran1,Berkowitz,Hann,Heller,Tuckerman,Lyubartsev,Hyvonen},
describing membranes with chemical accuracy, are limited to short
times and short lengths. Usually, atomistic models on times
exceeding microseconds and lengths larger than nanometers are not
practical \cite{Muller2}, especially when the equilibrium
structures of large systems are considered.

One of the strategy to overcome this problem is to group atoms into
effective particles that interact via effective potentials within so-called
coarse-grained models \cite{Theodorou}. Coarse-graining can reduce
considerably the degrees of freedom and the configurational space such that
larger systems and longer times can be reached. Monte-Carlo (MC) simulations
of phospholipid membranes with coarse-grained models of phospholipids \cite%
{Lipowsky,Klein1,Klein2,Muller4,Marrink,Deserno} are proved to capture the
essential properties of self-assembly of phospholipids into membranes, their
molecular structure and collective phenomena.

However, if the objective of the research is the equilibrium
properties and the molecular details of the self-assembled
structures, the equilibration time may largely exceed microseconds
and thus, may not be reached with MD simulations in a reasonable
time. Alternative strategy is to reach directly the minimum of the
free energy using methods based on mean field theories, thus,
avoiding computationally expensive equilibration process used in
MC and MD simulations.

The method of the Single Chain Mean Field (SCMF)
theory \cite{Ben-Shaul1,Ben-Shaul2,Szleifer,Bonet1,Bonet2,Szleifer2} is
particularly suitable for such purposes: it combines analytical theory of
the mean field type with the conformational variability accessible in MC
simulations. Thus, the accuracy of the method in describing the molecular
details of equilibrium structures can compete with MC simulations while the
speed of obtaining the results depends only on the speed of the numerical
solution of equations, thus it can be much faster than MC simulations.

The advantage of this method is the speed in localization of the
free energy minimum and the equilibrium properties as well as the
precision in the measurement of the equilibrium free energy which
is not straightforwardly accessible with MD and MC simulations.
The disadvantage is the mean-field nature of the solution that
does not include fluctuations and inter-particle correlations.
Nevertheless, the shortcomings of the approach can be compensated
by the combination of the SCMF theory with MD simulations which
can complement the method. Fast mean field method provides for
equilibrium structure with molecular details which, in turn, can
be used as input initial configuration for MD simulations and
provide the lacking dynamic information and fluctuations.

The SCMF theory, originally developed for the micellization
problem of low-molecular surfactants \cite{Ben-Shaul1,Ben-Shaul2},
describes a single molecule in the molecular fields. The
multichain problem is reduced to a single chain in the external
self-consistent field problem. The position, orientation and
configuration of the molecule depend on the surrounding fields
while the fields depend on the configurations of the molecules,
and finally the equilibrium fields are found self-consistently by
numerical solution of a system of nonlinear equations. This method
gives a detailed microscopic information on the configurations and
averaged positions of the molecules, the optimal shape and
structure of self-assembled structures, the distribution of
molecules in the aggregates, the critical micellar concentrations
as well as the optimal aggregation number and the size
distributions of the micelles and the aggregates
\cite{Bonet1,Bonet2}. This method is quite universal: it can be
applied to solutions of linear or branched polymers, solutions of
low-molecular weight surfactants and various additives, mixtures
of various components and structural and shape transitions.
However, the limiting factor restraining the use of the SCMF
theory is the computational realization of a stable code that can
efficiently solve self-consistent equations. There are no standard
packages with universal implementation of the SCMF theory, which
could make the use of the method accessible for a large number of
people like Mesodyn package \cite{Altevogt,Fraaije} for mesoscopic
modeling of phase separation dynamics or an integrated simulation
system for soft materials (OCTA) for multiscale modeling
\cite{Octa1,Octa2,Octa3}. Most of the previous works on SCMF
theory \cite{Ben-Shaul1,Ben-Shaul2,Szleifer,Bonet1,Bonet2} use
iterative methods or external mathematical libraries, which are
not optimized for a given problem and thus, computationally
unstable and highly demanding in computer resources, especially in
RAM memory. Thus, problems involving long molecules, large systems
and geometries more than 1D confront with the limit of available
computer resources and thus, making the solution of the problem
without special technical skills to be a quite difficult task. A
modification of the method, Single Chain \textit{in} Mean Field
calculations \cite{MullerSCF2,MullerSCF1,MullerSCF3}, avoids the
necessity to solve self-consistent equations. The direct solution
of the self-consistent equations is replaced by MC equilibration
of the chains in the \textit{quasi-instantaneous} fields
maintained at the most recent values and updated after a
predetermined number of MC moves. In practice, the MC
equilibration would slow down the calculation with respect to
direct solution of the equations, but it is still much faster than
the direct MC method.

It is noteworthy that SCMF method is similar in spirit to one of
the first mean field model of the phospholipid membranes by
Marcelja \cite{MarceljaNature,Marcelja1} constructed for modeling
of the fluid -- gel phase transition and based on phenomenological
potential of the Maier-Saupe type between the segments of the
tails. The boundary between the solvent and the membrane is
modeled as a planar surface with the phospholipid tails attached
with a fixed grafting density. Each tail interacts with the
neighbors via mean field which is found self-consistently.

In this work we report a computational tool providing relatively fast and
stable solution of the equations of the SCMF theory in different geometries
and different molecule structures. The SCMF theory is applied to model
phospholipid membranes. We show that three different coarse-grained models
of phospholipids can adequately describe the equilibrium properties and the
molecular structure of phospholipid membranes.

The paper is organized as follows. Theoretical principles of the
SCMF theory and main equations of the method are introduced in the
section \ref{sectheory} in the most general way. This theory is
then applied to the simulation of phospholipid bilayers and three
coarse-grained models of phospholipid molecule are introduced in
section \ref{secapplic}. These models and the resulting
equilibrium phospholipid bilayers are compared with experimental
data and between each other in section \ref{secresults}. Last
section \ref{secsconclusions} summarizes the obtained results
while the computational details which are necessary for the
implementation of the method are described in the Appendix
\ref{seccompdet}.

\section{Theory \label{sectheory}}

The SCMF theory is an example of the Self-Consistent Field (SCF) method
where a single chain is described at the molecular level while the
interactions between different chains are described through a mean molecular
field which is found self-consistently.

The conformations of a single chain are generated with the Rosenbluth
algorithm \cite{Rosenbluth} or MC simulations \cite{Theodorou,FrenkelBook}
and the intra-molecular interactions are calculated exactly using the model
potential for interactions between the segments of the chain (see Appendix %
\ref{secgeneration} for details). The probabilities of individual chain
conformations depend on the mean molecular fields, while the values of the
fields are calculated as the average properties of individual conformations.
The resulting equations for the mean fields and the probabilities of the
conformations are solved self-consistently with the original method
described in the Appendix \ref{secsolution}. The solution of these equations
gives the equilibrium structures and the concentrations profiles of all
components in the system as well as the most probable conformations of
individual molecules. The power of this method is the speed in obtaining
solutions (faster than MC and much faster than MD simulations) and the
precise calculation of the energies.

This method is quite universal and can be applied for the mixture of an
arbitrary number of molecules of different types interacting with each other
through the mean fields. The free energy of a system containing $%
N_{1},N_{2},...,N_{M}$ molecules of types $1,2,...,M$ can be written as a
sum of three terms, the entropy, intra- and inter-molecular terms, all
written in terms of $kT$,
\begin{equation}
F=-\left\langle S\right\rangle +\left\langle H^{intra}\right\rangle
+\left\langle H^{inter}\right\rangle  \label{FreeEnergyGeneral}
\end{equation}%
where the angular brackets denote the average over the probability
distribution function (pdf) $\rho $ of the system, $\left\langle \ldots
\right\rangle =\int \ldots \rho d\rho $. This function contains all
information about the equilibrium state of the whole system. If there are no
strong correlations between interacting particles, such as ion pair
formation or bonds formation, we can neglect the correlations between the
molecules and use the mean field approximation: the pdf $\rho $ of many
component system factorizes into the product of single molecules pdfs $\rho
_{\alpha }(\Gamma _{\alpha i})$,
\begin{equation}
\rho \approx \prod_{\alpha =1}^{M}\prod_{i=1}^{N_{\alpha }}\rho _{\alpha
}(\Gamma _{\alpha i})  \label{ProbabilitySplit}
\end{equation}%
where $\Gamma _{\alpha i}$ is the conformation of $i$-th molecule of type $%
\alpha $. Such factorization into contributions of individual molecules
allows us to derive a close set of equations for $\rho _{\alpha }(\Gamma
_{\alpha i})$.

The entropy term in the expression (\ref{FreeEnergyGeneral}) is written as
\begin{equation}
\left\langle S\right\rangle =-\left\langle \ln \rho \Lambda \right\rangle
,\quad  \label{Entropy_Pre}
\end{equation}%
where the factor $\Lambda $ is the de Broglie length which has a quantum
mechanics origin,

\begin{equation}
\Lambda =\prod_{\alpha =1}^{M}\Lambda _{\alpha }^{N_{\alpha }}N_{\alpha
}!\approx \prod_{\alpha =1}^{M}\left( \frac{N_{\alpha }\Lambda _{\alpha }}{e}%
\right) ^{N_{\alpha }}
\end{equation}%
Since the constants $\Lambda _{\alpha }$ do not appear in the final
expressions, they can be treated as unknown normalization constants. Thus,
after factorizing the pdf $\rho $ (\ref{ProbabilitySplit}), the entropy term
(\ref{Entropy_Pre}) reads
\begin{equation}
\left\langle S\right\rangle \approx -\sum_{\alpha =1}^{M}N_{\alpha
}\left\langle \ln \frac{\rho _{\alpha }N_{\alpha }}{e}\Lambda _{\alpha
}\right\rangle  \label{Entropy}
\end{equation}%
where the brackets on the right hand side denote the average over the single
molecule pdf $\rho _{\alpha }$. Similar arguments allow us to write the
intra-molecular energy of the system as a sum of contributions of the single
molecules of different types
\begin{equation}
\left\langle H^{intra}\right\rangle \approx \sum_{\alpha =1}^{M}N_{\alpha
}\left\langle H_{\alpha }^{intra}\right\rangle  \label{Intramolecular_Energy}
\end{equation}

Before writing a similar expression for the inter-molecular part, we assume
that molecules of each type $\alpha $ comprise of subunits with different
chemical structure, and thus different energy parameters. Subunits can
represent Kuhn segments in the chain or, in case of coarse-graining
description, the groups of atoms or beads in a coarse-grained model. Thus,
the interaction energy between a molecule of type $\alpha $ in the
conformation state $\Gamma _{\alpha }$ and a molecule of type $\beta $ in
the conformation state $\Gamma _{\beta }$, can be written as a sum over the
types of beads $a$ as
\begin{equation}
H_{\alpha \beta }^{inter}\left( \Gamma _{\alpha },\Gamma _{\beta }\right)
=\sum_{a}\int u_{\alpha }^{a}(\Gamma _{\alpha },\pmb r)c_{\beta }^{a}(\Gamma
_{\beta },\pmb r)d\pmb r  \label{HIntermolecularPair}
\end{equation}%
In this expression $c_{\beta }^{a}(\Gamma _{\beta },\pmb r)$ is the
concentration of units of type $a$ at the point $\pmb r$ of a molecule of
type $\beta $ in the conformation state $\Gamma _{\beta }$. These units
interact with the field $u_{\alpha }^{a}(\Gamma _{\alpha },\pmb r)$ created
by the molecules of type $\alpha $. Thus, factorizing the pdf $\rho $ (\ref%
{ProbabilitySplit}) one can write the inter-molecular interaction free
energy in the form
\begin{equation}
\left\langle H^{inter}\right\rangle \approx \frac{1}{2}\sum_{\alpha ,\beta
=1}^{M}N_{\alpha }(N_{\beta }-\delta _{\alpha \beta })\sum_{a}\int
\left\langle u_{\alpha }^{a}(\pmb r)\right\rangle \left\langle c_{\beta
}^{a}(\pmb r)\right\rangle d\pmb r  \label{Intermolecular_Energy}
\end{equation}%
where $\delta _{\alpha \beta }$ is the delta symbol. Thus, the interaction
free energy is represented by the interaction of the average concentration
of beads with the average fields at each point.

The free energy is usually coupled with the incompressibility condition
implying that the sum of the concentrations of all components in a solution
is fixed. However, this condition implies the hard core repulsion between
all beads in the system, and thus creates computational problems in
converging the SCMF equations. To overcome this problem, we introduce the
explicit incompressibility condition in every point $\pmb r$ in the system,
\begin{equation}
\sum_{\alpha =1}^{M}N_{\alpha }\left\langle \phi _{\alpha }(\pmb %
r)\right\rangle =\phi _{0}  \label{IncompressibilityConstrain}
\end{equation}%
where $\left\langle \phi _{\alpha }(\pmb r)\right\rangle $ is the average
volume fraction occupied by a molecule of type $\alpha $ in the point $\pmb %
r $, while $\phi _{0}$ is the total volume fraction occupied by the
molecules of all types.

Combining all terms (\ref{Entropy}), (\ref{Intramolecular_Energy}), (\ref%
{Intermolecular_Energy}) together with the incompressibility condition (\ref%
{IncompressibilityConstrain}), the free energy of the system can be written
as
\begin{equation}
\begin{split}
&F[\rho _{1},\rho _{2},...,\rho _{M}] \approx \sum_{\alpha
=1}^{M}N_{\alpha }\left\langle \ln \frac{\rho _{\alpha }N_{\alpha
}}{e}\Lambda _{\alpha }\right\rangle +\\
&\sum_{\alpha =1}^{M}N_{\alpha }\left\langle H_{\alpha
}^{intra}\right\rangle + \\
& \frac{1}{2}\sum_{\alpha ,\beta =1}^{M}N_{\alpha }(N_{\beta
}-\delta
_{\alpha \beta })\sum_{a}\int \left\langle u_{\alpha }^{a}(\pmb %
r)\right\rangle \left\langle c_{\beta }^{a}(\pmb r)\right\rangle d\pmb r+ \\
& \int \lambda (\pmb r)\left( \phi _{0}-\sum_{\alpha
=1}^{M}N_{\alpha }\left\langle \phi _{\alpha }(\pmb
r)\right\rangle \right) d\pmb r
\end{split}
\label{FreeEnergyFull}
\end{equation}%
where $\lambda (\pmb r)$ is a Lagrange multiplier and all the averages are
taken over the single-molecule pdfs $\rho _{\alpha }$. Minimization of this
functional with respect to $\rho _{\alpha }$ gives the pdf of a single
molecule of type $\alpha $
\begin{eqnarray}
&&\rho _{\alpha }(\Gamma _{\alpha }) =\frac{1}{Z_{\alpha }}\exp
\left( -H_{\alpha }^{intra}(\Gamma _{\alpha })-\sum_{\beta
=1}^{M}(N_{\beta
}-\delta _{\alpha \beta })\times \right.  \notag \\
&&\left. \sum_{a}\int u_{\alpha }^{a}(\Gamma _{\alpha },\pmb r)\left\langle
c_{\beta }^{a}(\pmb r)\right\rangle d\pmb r+\int \lambda (\pmb r)\phi
_{\alpha }(\Gamma _{\alpha },\pmb r)d\pmb r\right)
\label{ProbabilityExpression}
\end{eqnarray}%
where $Z_{\alpha }$ is a normalization constant which is found from the
normalization condition $\int \rho _{\alpha }(\Gamma _{\alpha })d\Gamma
_{\alpha }=1$. This constant has a meaning of a partition function of the
system at equilibrium and $-\ln Z_{a}$ is the total free energy of the
system at equilibrium.

Once the average concentrations $\left\langle c_{\beta }^{a}(\pmb %
r)\right\rangle $ and the volume fractions $\left\langle \phi _{\beta }(\pmb %
r)\right\rangle $ are known, this expression allows to calculate the
probabilities of each conformation $\rho _{\alpha }(\Gamma _{\alpha })$ for
all molecules in the system. In turn, if the probabilities $\rho _{\alpha
}(\Gamma _{\alpha })$ are known, the average concentrations and volume
fractions, being the the molecular fields in this problem are found from the
self-consistency conditions,
\begin{eqnarray}
\left\langle c_{\alpha }^{a}(\pmb r)\right\rangle &=&\int c_{\alpha
}^{a}(\Gamma _{\alpha },\pmb r)\rho _{\alpha }(\Gamma _{\alpha })d\Gamma
_{\alpha }  \label{MainEquations} \\
\left\langle \phi _{\alpha }(\pmb r)\right\rangle &=&\int \phi _{\alpha
}(\Gamma _{\alpha },\pmb r)\rho _{\alpha }(\Gamma _{\alpha })d\Gamma
_{\alpha },  \notag
\end{eqnarray}%
representing the averages over the probabilities of conformations.

The probability of each conformation can be written as $\rho _{\alpha
}(\Gamma _{\alpha })=\frac{1}{Z_{\alpha }}\exp [-H_{eff}(\Gamma _{\alpha })]$%
, where $H_{eff}(\Gamma _{\alpha })$ is the effective Hamiltonian given by (%
\ref{ProbabilityExpression}), which describes the system at equilibrium. The
last term in the effective Hamiltonian corresponds to the steric repulsion
of beads of all types. If one of the components is a one bead solvent, the
only degree of freedom of the solvent molecules is the position in space, $%
\pmb r$. Hence, the volume fraction of the solvent, can be found from the
incompressibility condition (\ref{IncompressibilityConstrain})
\begin{equation}
\phi _{s}(\pmb r)=\phi _{0}-\sum_{\alpha =1,\alpha \neq s}^{M}N_{\alpha
}\left\langle \phi _{\alpha }(\pmb r)\right\rangle  \label{FractionOfSolvent}
\end{equation}

In addition, the Lagrange multiplier $\lambda (\pmb r)$ can be expressed
through the concentration of the solvent. The pdf of the solvent $\rho _{s}(%
\pmb r)$, the concentration $c_{s}(\pmb r)$ and the volume fraction occupied
by the solvent $\phi _{s}(\pmb r)$ are related via the following expressions
\begin{equation}
\rho _{s}(\pmb r)=\frac{c_{s}(\pmb r)}{N_{s}}=\frac{\phi _{s}(\pmb r)}{%
v_{s}N_{s}}\quad  \label{ExpressionsForSolvent}
\end{equation}%
where $N_{s}$ is the number and $v_{s}$ is the volume of the solvent
molecule, while the molecular field

\begin{equation}
\phi _{s}(\pmb r,\pmb r^{\prime })\approx v_{s}\delta (\pmb r-\pmb
r^{\prime }) \label{abb}
\end{equation}%
where $\delta (\pmb r-\pmb r^{\prime })$ is the Dirac
delta-function. Substitution of (\ref{ExpressionsForSolvent}) and
(\ref{abb}) into (\ref{ProbabilityExpression}) gives the
approximate expression for the Lagrange multiplier $\lambda (\pmb
r)$
\begin{equation}
v_{s}\lambda (\pmb r)\approx \ln \phi _{s}(\pmb r)+\sum_{\beta
=1}^{M}(N_{\beta }-\delta _{s\beta })\sum_{a}\int u_{s}^{a}(\pmb r,\pmb %
r^{\prime })\left\langle c_{\beta }^{a}(\pmb r^{\prime })\right\rangle d\pmb %
r^{\prime }  \label{LagrangeMultiplier}
\end{equation}%
where we have assumed that $H_{s}^{intra}(\pmb r)=0$ and omitted few
constants, which will cancel out by the normalization of $\rho_\alpha $.

It is noteworthy, that in our equations the fields $\left\langle
\phi _{\alpha }(\pmb r)\right\rangle $ and $\left\langle c_{\alpha }^{a}(%
\pmb r)\right\rangle $ formally are not related although they both
correspond to the concentration of monomers. They are indeed
related through the volume of the monomers only if the molecules
are composed of \textit{non-overlapping} beads, when the distance
between the centers is larger than the the diameter. However, if
we want to conserve the possibility of describing the overlapping
beads, the coefficient of proportionality between $\left\langle
\phi _{\alpha }(\pmb r)\right\rangle $ and $\left\langle c_{\alpha }^{a}(%
\pmb r)\right\rangle $ is not known a priori and we keep them as
independent variables.

The equations (\ref{ProbabilityExpression}),(\ref{MainEquations}),(\ref%
{FractionOfSolvent}) and (\ref{LagrangeMultiplier}) form a closed set of
non-linear equations of the SCMF theory. The solution of these equations
gives the equilibrium structures, the self-consistent molecular fields such
as the concentration profiles of the beads of each type and the distribution
of the solvent, and the probabilities of each conformation of the molecules
in the fields and the accurate measure of equilibrium free energies. The
implementation of the computational method as well as the technical details
related to the solution of the equations are present in Appendix \ref%
{seccompdet}.

\section{Application to phospholipid membranes \label{secapplic}}

The method of the SCMF theory described so far is applied to model the
equilibrium structures and properties of phospholipid membranes. The SCMF
theory can provide the detailed information on the microscopic structure of
the phospholipid layer such as concentration profiles of all groups of atoms
of the phospholipid molecule, the thickness of the membrane, the average
area per phospholipid head group and the mechanical properties such as
compressibility of the membrane and the surface tension. All these
parameters can be measured in the experiment \cite{Nagle1,Nagle2,Nagle3} and
thus, allowing the direct comparison with experimental data.

We have developed the C++ code which runs in parallel using OpenMP shared
memory platform on the 32-core AMD machines (see Appendix \ref{secsolution}%
). To simulate a flat phospholipid layer we use one dimensional geometry and
discretize the space into parallel cells with the same value of the average
fields. The periodic boundary condition is applied together with the
assumption of the symmetry of the layer with respect to the central plane.

\begin{figure}[t]
\begin{center}
\includegraphics[width=8.3cm]{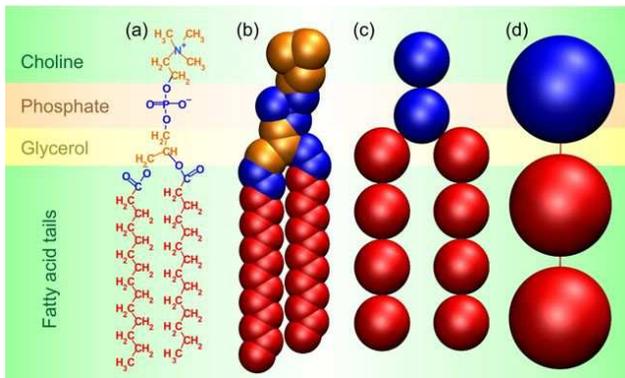}
\end{center}
\caption{(a) Chemical structure of the DMPC phospholipid molecule.
44-beads (b), 10-beads (c) and 3-beads (d) models of phospholipid
molecules used in the present calculations. Blue beads correspond
to hydrophilic monomers, red beads correspond to hydrocarbon
monomers in the tails and orange beads correspond to hydrophobic
monomers in the heads. The interaction parameters of red and
orange beads are the same.} \label{Figure_LipidModels}
\end{figure}

Despite the complexity of the structure of the phospholipid
membranes, the driving force for the self-assembly of
phospholipids into bilayers is the amphiphile nature of
phospholipids. That is why, we only consider the units of two
types, hydrophilic, H, and hydrophobic, T. The units of both types
interacts with each other through the square well potentials. We
assume that all units, T and H, and the solvent molecules S, have
the same size and the same interaction range, while the
interaction energies are different. We consider three types of
interactions, interactions between hydrophobic units, T-T,
interactions between hydrophobic units and solvent, T-S and
hydrophilic units with solvent, H-S. In the following we will show
that such assumptions are justified for quantitative study of
phospholipid layers and, in principle, cohere with approximations
used in successful models of phospholipid membranes (see e.g. Ref.
\citenum{Deserno}). However, if necessary, more types of beads and
more complicated potentials can be implemented.

\begin{table}
\caption{Parameters used for the simulation of phospholipid
bilayer for three models of DMPC phospholipid molecule.}
\label{Table1}{\small \
\begin{tabular}{|l|r|r|r|}
\hline
& 44B & 10B & 3B \\ \hline\hline
Units radius (\AA ) & $1.90$ & $2.50$ & $4.05$ \\ \hline
Interaction range (\AA ) & $5.70$ & $7.50$ & $12.15$ \\ \hline
T-T contact energy ($kT$) & $-0.40$ & $-1.50$ & $-2.10$ \\ \hline
T-S contact energy ($kT$) & $0.00$ & $0.00$ & $0.00$ \\ \hline
H-S contact energy ($kT$) & $-0.10$ & $-0.20$ & $-0.15$ \\ \hline
Bond length (\AA ) & $1.5$ & $5.0$ & $10.0$ \\ \hline
Occupied volume fraction $\phi _{0}$ & $0.700$ & $0.675$ & $0.675$ \\ \hline
Sampling (number of configurations) & $3\times 10^{6}$ & $10^{6}$ & $10^{6}$
\\ \hline
Simulation box size (\AA ) & \multicolumn{3}{c|}{$120.0\times 120.0\times
62.7$} \\ \hline
\end{tabular}
}
\end{table}

The chemical structure of the DMPC molecule is depicted in Figure \ref%
{Figure_LipidModels}a. We present three models of DMPC phospholipid molecule
at a different level of coarse-graining.The first, most detailed model,
represents the DMPC phospholipid as a two tails molecule of 44 beads (Figure %
\ref{Figure_LipidModels}b). A carbon group \ce{CH2} of the
molecule is assigned to one T bead. A phosphate group, \ce{PO4},
is represented by 5 H overlapping beads, placed at the vertexes
and center of a tetrahedron. The choline group \ce{NC4H11} is
represented in a similar way: one H bead in the center of a
tetrahedron is surrounded by 4 T beads placed in the corners. The
\ce{COO} groups are represented by 2 H beads. The angles between
most of the bonds are fixed at value $120^{\circ }$. The torsion
angles the sequence of groups \ce{CH2-CH2-CH2} are allowed to have
only three fixed values, $0^{\circ }$, $120^{\circ }$ and
$240^{\circ }$, which corresponds to cis- and trans- conformations
of the groups. The parameters of the model (Table \ref{Table1})
were adjusted with a series of fast simulations (conformational
sampling size $\sim 30-100$ thousands and $40$ layers in the
simulation box) while the accurate data were obtained with several
millions of conformations and hundred layers in the box.

Two others, more coarse-grained and less detailed, models are depicted in
Figures \ref{Figure_LipidModels}c and \ref{Figure_LipidModels}d. One of them
represents two tails phospholipid with 10 beads and another is simply
3-beads freely joined together. In contrast to 44-beads model, the
conformational space of these models is much more restricted and we need
less sampling to produce accurate results. Thus, the simplest 3-beads model
is the fastest and demanding less computer resources phospholipid model
which can simulate the self-assembly of phospholipids into the bilayers with
realistic properties of DMPC phospholipid membranes.

\begin{figure*}[tbp]
\begin{center}
\includegraphics[width=15cm]{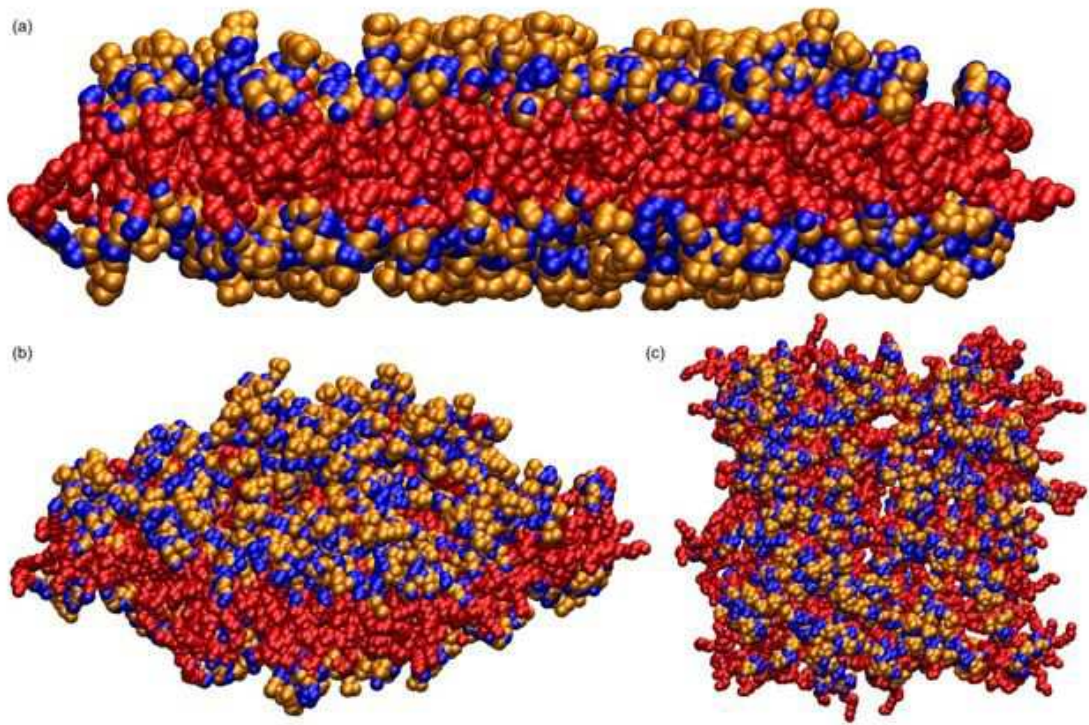}
\end{center}
\caption{Typical mean field "snapshots" (set of most-probable
conformations of the lipids) of the phospholipid membrane within
the 44-beads model.} \label{Figure_Membrane}
\end{figure*}

It is important to note, that we simulate a finite size box with the fixed
volume $V$ and the fixed number of phospholipid molecules $N_{l}$ with the
free energy $F$ given by the expression (\ref{FreeEnergyFull}). However, we
can extrapolate our data to much larger system with larger number of
phospholipids $N_{l}^{\ast }$ and larger number of solvent molecules. The
description of a larger system is possible, if we do not take into account
any energy contributions related to the perpendicular undulations and
bending of the membrane. In this case the simulation box represent a
self-similar part of a larger system. The free energy of the larger system $%
F^{\ast }$ is in a simple relation with the free energy per lipid $f$
\begin{equation}
F^{\ast }\approx {const}+N_{l}^{\ast }f  \label{FreeEnergyTransformation}
\end{equation}%
which, in turn, is related to the free energy of the simulation box minus
the entropy of the solvent

\begin{equation}
fN_{l}=F-V\frac{\phi _{0}}{v_{s}}\ln \frac{\phi _{0}}{v_{s}}
\end{equation}%
where $\phi _{0}/v_{s}$ is the concentration of the pure solvent.

Thus, the calculation of free energy of the box $F$ with different number of
molecules in the simulation box gives the free energy of the membrane per
lipid $f$. The minimum of the free energy per lipid $f$ corresponds to the
equilibrium density of the membrane, thus the second derivative of this
energy with respect to the area per lipid in the minimum gives the
compressibility modulus of the layer

\begin{equation}
K=2A_{eq}f^{\prime \prime }(A_{eq})  \label{K}
\end{equation}
representing a measure of the rigidity of the membrane.

\section{Results and discussion \label{secresults}}

The numerical implementation of the method in the form of two structurally
independent modules: the generation of a single molecule conformations and
the solution of equations (see Appendix \ref{seccompdet}), permits us to
consider several models of the molecule at the same time. Thus, we can
compare our three models of phospholipids and test their performance in
reproducing the thermodynamic properties of phospholipid membranes.

The most detailed 44-beads model (Figure \ref{Figure_LipidModels}b) results
in equilibrium structures of membranes where the phospholipids are assembled
into flat bilayers in fluid phase. Since the output of the method is the
probabilities of each molecular conformation in the self-consistent fields,
the most probable conformations may be used for the visualization of the
resulting structures which correspond to the solutions of the equations.
Typical mean field "snapshots" of the equilibrium bilayer structures
obtained with the 44-beads model are present at Figure \ref{Figure_Membrane}%
. Although the picture visually resemble instant snapshots of MC
simulations, this is not a representation of interacting molecules at a
given moment of time, but these are the most probable conformations of a
single chain in the fields corresponding to the equilibrium solution. Such
"snapshots" are helpful for visualization purposes in order to distinguish
between different molecular structures.

The equations of the SCMF theory may result in several sets of
solutions corresponding to various equilibrium and even metastable
structures. To this end, the numerical method is implemented in
such a way that simultaneously several solutions corresponding to
local and global minima of the free energy can be found. The
obtained solutions can then be ranged by their free energy or by
the concentration profile. Even in the restricted 1D geometry of
flat phospholipid bilayers with periodic boundary condition the
method finds several self-assembled structures: homogeneous
solution of lipids, one single layer, formed either at the center
or at the edge of the box and two or even three layers in the box.
Each structure has its own free energy which allows to determine
the equilibrium properties.

\begin{figure}[tbp]
\begin{center}
\includegraphics[width=8.3cm]{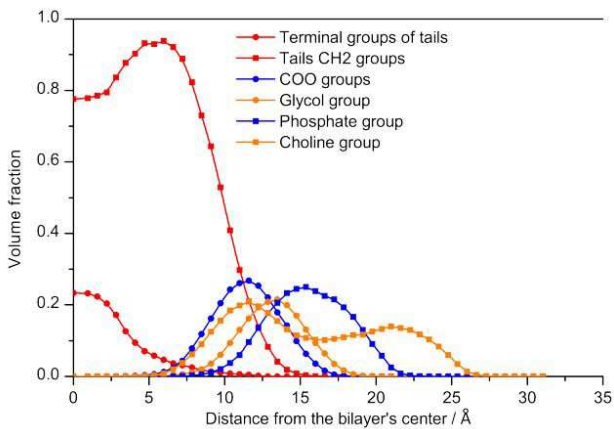}
\end{center}
\caption{Detailed equilibrium concentration profile of the
phospholipid bilayer obtained with the 44-beads model.}
\label{Figure-Profile-Detailed}
\end{figure}

The detailed microscopic information concerning the mechanic and
thermodynamic properties of the simulated flat bilayers can be compared
directly against the experimental data for phospholipid bilayers \cite%
{Nagle1,Nagle2,Nagle3}. Thermodynamic properties of the membranes obtained
from the measurements are the thickness of the membrane (determined as a
distance between the midpoints of the outer slopes of the H beads
concentration profiles), the full thickness of the bilayer region
(determined as a thickness of the region where the total concentration of
the lipid molecules is larger than zero), the thickness of the hydrophobic
core (measured as a distance between the midpoints of the slopes of the T
beads concentration profiles), the distances between different groups of the
phospholipid molecule, membrane rigidity measured by the compressibility
modulus and the interfacial area per lipid molecule. Thus, the
coarse-grained parameters of each model of phospholipid molecule are chosen
in such a way that all these microscopic properties correspond to the
experimental values of the DMPC phospholipid bilayers. The parameters of
these three models are summarized in Table \ref{Table1}.

\begin{table*}
\caption{Comparison of the equilibrium properties of phospholipid
bilayer obtained with three lipid models with experimental data
and full atomistic MD simulations of the DMPC lipid
bilayer.}{\small \
\begin{tabular}{|l|r|r|r|r|}
\hline & 44B & 10B & 3B & DMPC \\ \hline\hline
Membrane thickness (\AA ) & 35 & 36 & 45 & $44.2$\textsuperscript{\emph{a}} \\
\hline Full thickness of the bilayer region (\AA ) & 51 & 44 & 56
& $53$\textsuperscript{\emph{b}} \\ \hline Thickness of
hydrophobic core (\AA ) & 24 & 24 & 28 &
$26.2$\textsuperscript{\emph{a}} \\ \hline
Distance between heads (\AA ) & 30 & 28 & 36 & $36$\textsuperscript{\emph{a}} \\
\hline Distance between phosphate groups (\AA ) & 30 & - & - &
$32$\textsuperscript{\emph{b}} \\ \hline
Distance between choline groups (1st peak) (\AA ) & 22 & - & - & $37$\textsuperscript{\emph{b,d}} \\
\hline
Distance between choline groups (2nd peak) (\AA ) & 42 & - & - & $37$\textsuperscript{\emph{b,d}} \\
\hline Distance between glycol groups (\AA ) & 26 & - & - &
$27$\textsuperscript{\emph{b}}
\\ \hline Distance between \ce{COO} groups (\AA ) & 24 & - & - &
$25$\textsuperscript{\emph{b}} \\ \hline Thickness of terminal
groups region (\AA ) & 6 & - & - & $10$\textsuperscript{\emph{b}}
\\ \hline Interfacial area per lipid (\AA $^{2}$) & $70\pm 5$ &
$65\pm 11$ & $60\pm 2$ & $59.6$\textsuperscript{\emph{a}} \\
\hline Compress. constant
(dyn/cm) & $135\pm 11$ & $560\pm 120$ & $269\pm 11$ & $257$\textsuperscript{\emph{c}} \\
\hline
\end{tabular}

\textsuperscript{\emph{a}} Experimental data by
\citeauthor{Nagle2}\cite{Nagle2}

\textsuperscript{\emph{b}} Calculated from MD simulations data by
\citeauthor{Damodaran1} \cite{Damodaran1}

\textsuperscript{\emph{c}} Experimental data by
\citeauthor{Nagle3}\cite{Nagle3}

\textsuperscript{\emph{d}} Average distance between choline
groups. } \label{Table2}
\end{table*}

The most detailed 44-beads model provides all information about the
composition of the membrane as well as the position of different groups of
the molecule in the bilayer. The concentration profile of the bilayer at
equilibrium is shown in Figure \ref{Figure-Profile-Detailed}. The profiles
are normalized to unity by division by the occupied volume fraction $\phi
_{0}$. One can see, that the terminal groups of the lipid tails are situated
in the center of the layer, which is surrounded by the area filled with
other carbon groups of the tails. The groups of hydrophilic units
representing the \textquotedblleft heads\textquotedblright\ are in the
surface layer. Such position of the groups is in agreement with the
experimental data for such system \cite{Nagle2}. It is noteworthy, that the
concentration profile of choline group has two peaks reflecting the position
of phospholipid heads in the layer: approximately half of the heads are
oriented perpendicular to the layer plane, while the other half is parallel
to the layer plane. This is the outcome of the model of the phospholipid
molecule and the parametrization of the interactions. This can certainly
be improved once the reliable experimental data or results of MD simulations
concerning the equilibrium position of the heads in the layer is available.

One of the advantages of the SCMF method is the accurate
measurements of the free energies of equilibrium structures. In
our model this corresponds to the free energy of formation of flat
phospholipid layer. It does not include the entropy of thermal
undulations in the perpendicular direction. In particular,
undulations may result in the broadening of the phospholipid
membrane. However, they become important on the scales much larger
than the size of the simulation box and, in principle, can be
included as an additional contribution to the free energy of the
layer.

\begin{figure}[tbp]
\begin{center}
\includegraphics[width=8.3cm]{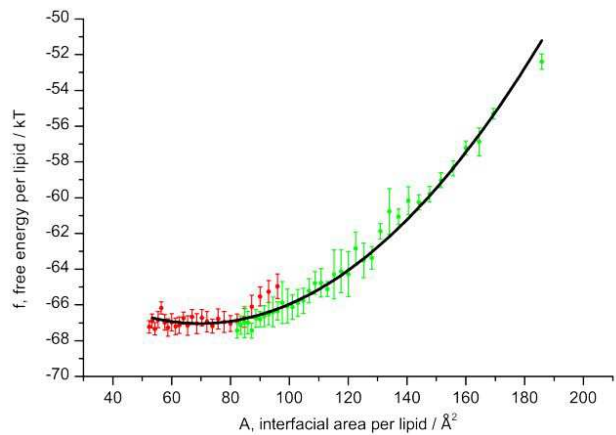}
\end{center}
\caption{Free energy per lipid $f$ vs. interfacial area per lipid
$A$ for the 44-beads lipid model, calculated from solutions with
one (red points) and two (green points) bilayers in the simulation
box. Parameters of the fitting curve $y=a_{0}+a_{1}x+a_{2}x^{2}$
are $a_{0}=-61.3\pm 0.5$, $a_{1}=-0.164\pm 0.010$,
$a_{2}=(1.17\pm0.04)\times 10^{-3}$.} \label{Figure-45B-Energy}
\end{figure}

The rigidity of the membrane can be obtained from the free energy of the
bilayer. Changing the number of lipids in the box, we get the free energy
per lipid molecule $f$ as a function of the interfacial area per lipid $A$
(Figure \ref{Figure-45B-Energy}). Although the error bars are quite large
because of the large conformational space with respect to the sampling used
for the calculations, the energy points can be quite good approximated with
the polynomial of the second order. This curve allows us to calculate the
compressibility of the layer with respect to compression and stretching in
the lateral direction. The second derivative is related to the
compressibility modulus defined in eq. (\ref{K}). The resulting
compressibility modulus, the thickness of the membrane, positions of various
groups and the area per lipid are shown in Table \ref{Table2}. The obtained
values agree quite well with the experimental data. However, if better
agreement is needed, the parameters of the models can be tuned by the
additional series of run of calculations with a large number of
conformations in the sampling.

The free energy per lipid $f$ (Figure \ref{Figure-45B-Energy}) has a minimum
at $A\approx 60$\AA $^{2}$, which corresponds to the experimental value of
the interfacial area per lipid in the bilayer in the fluid phase. Howether,
the values of $A$ below $50$\AA $^{2}$ cannot be obtained with the same set
of parameters, because this range corresponds to dense packing in the core
of the layer. Further decrease of interfacial area per lipid is accompanied
with structural rearrangements of the phospholipid environment and strong
correlations between the neighboring molecules. These correlations induce
phase transitions in lipid membranes \cite{Koynova}, the main transition to
the gel phase occurs at $A\approx 40$\AA $^{2}$. However, such correlations
are well beyond the limit of validity of the mean field theory, where the
movements of individual molecules are supposed to be uncorrelated (\ref%
{ProbabilitySplit}).

\begin{figure}[tbp]
\begin{center}
\includegraphics[width=8.3cm]{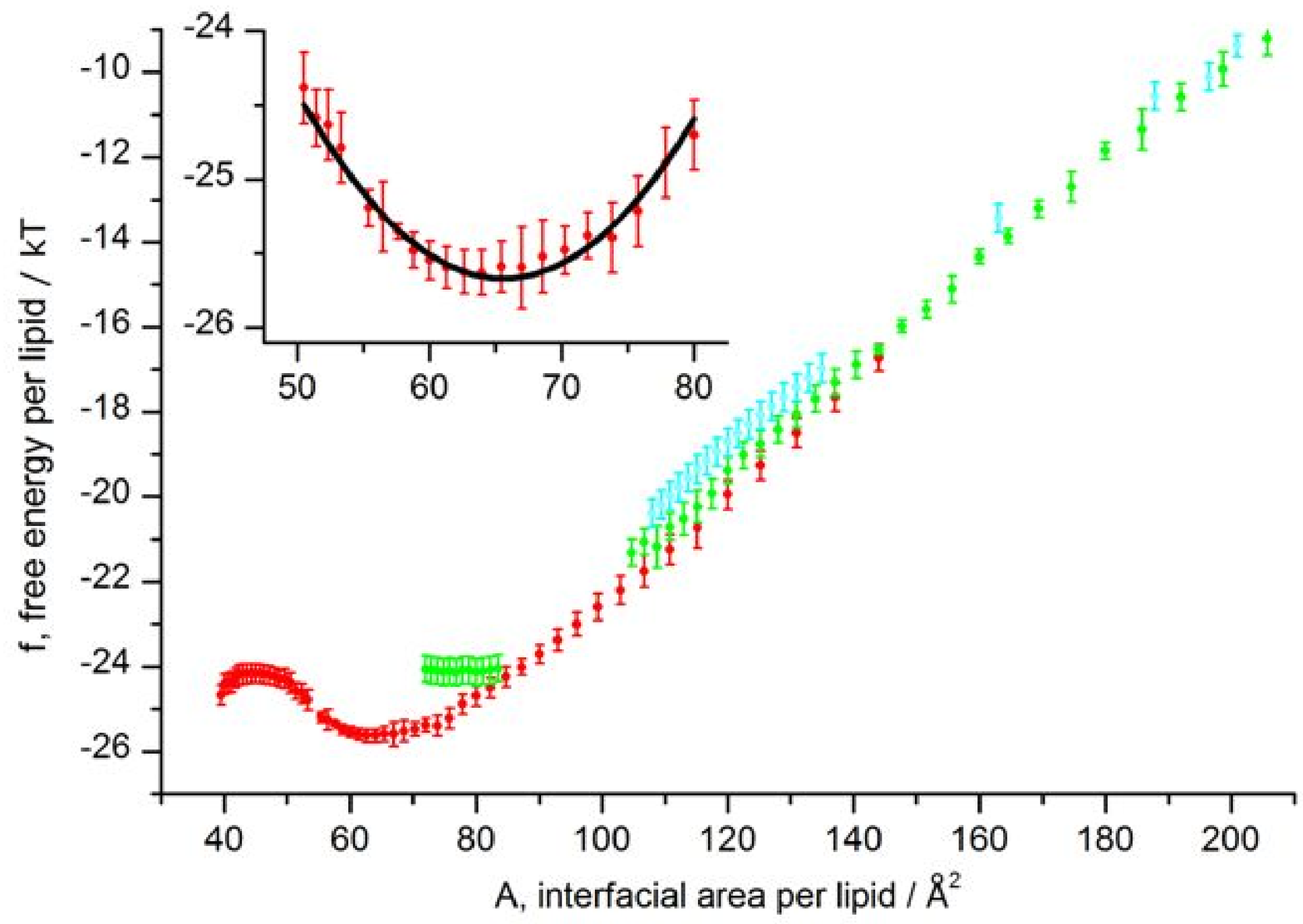}
\end{center}
\caption{Free energy per lipid $f$ vs. interfacial area per lipid
$A$ for the 10-beads lipid model, calculated from solutions with
one (red points), two (green points) and three (blue points)
bilayers in the simulation box. The inset represents the enlarged
region of the minimum. Parameters of the fitting curve
$y=a_{0}+a_{1}x+a_{2}x^{2}$ are $a_{0}=-3.0\pm3.0,$
$a_{1}=-0.68\pm0.08,$ $a_{2}=(5.2\pm0.6)\times 10^{-3}$.}
\label{Figure-10B-Energy}
\end{figure}

\begin{figure}[t]
\begin{center}
\includegraphics[width=8.3cm]{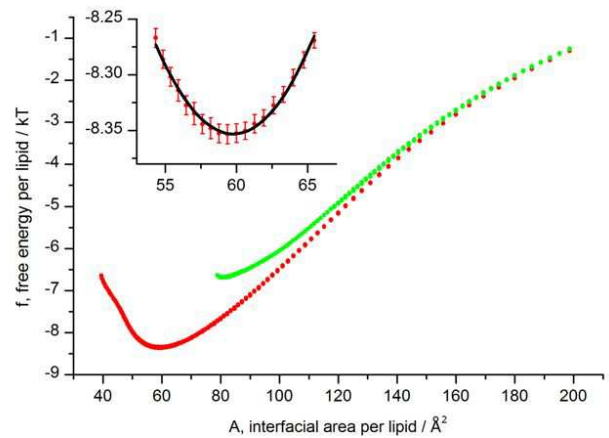}
\end{center}
\caption{Free energy per lipid $f$ vs. interfacial area per lipid
$A$ for the 3-beads lipid model, calculated from solutions with
one (red points) and two (green points) bilayers in the simulation
box. The inset represents the enlarged region of the minimum.
Parameters of the fitting curve $y=a_{0}+a_{1}x+a_{2}x^{2}$ are $
a_{0}=-1.4\pm0.2,$ $a_{1}=-0.325\pm0.008,$ $a_{2}=(2.72\pm
0.06)\times 10^{-3}$.} \label{Figure-3B-Energy}
\end{figure}

\begin{figure*}[tbp]
\begin{center}
\includegraphics[width=12cm]{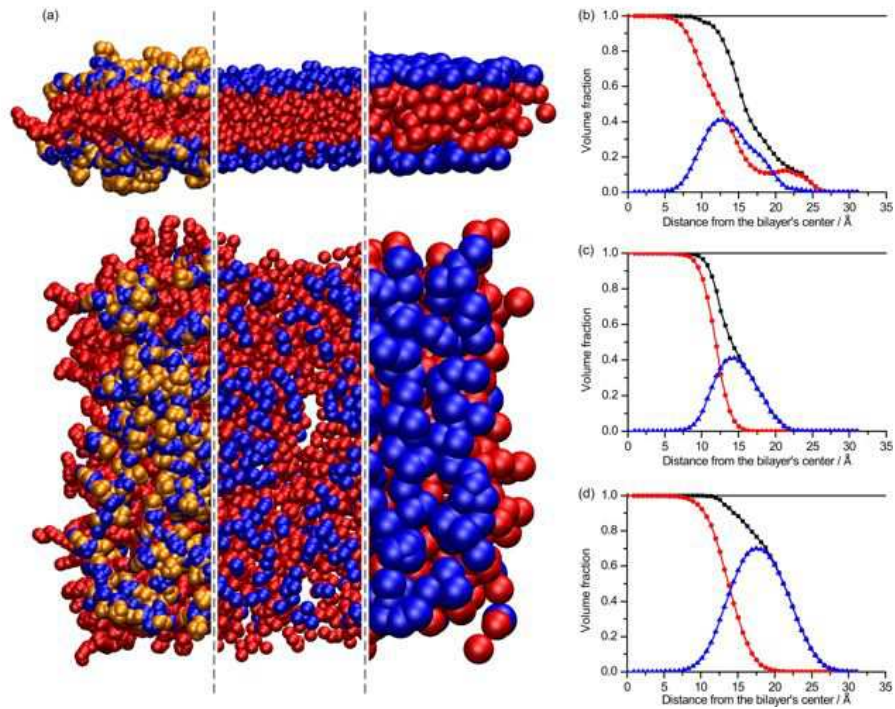}
\end{center}
\caption{Mean field equilibrium "snapshots" (set of most-probable
conformations of the lipids) and the concentration profiles of
tails (T, red lines), heads (H, blue lines) and the phospholipids
(T+H, black lines) obtained with 44-, 10- and 3-beads models.}
\label{Figure-Snaps}
\end{figure*}

Similar calculations carried out for the 10-beads model show better
reproducibility due to considerably reduced conformational space. In fact, 1
million of conformations is enough for calculations with different sampling
to almost coincide in one curve with very small error bars (Figure \ref%
{Figure-10B-Energy}). The absolute values of the free energy $f$
are shifted with respect to that of 44-units model, which is the
consequence of different sizes of the units and the interactions
parameters. However, the energy differences, the density profiles
and the compressibility modulus are very close to that obtained
with the previous model (Table \ref{Table2}). It is interesting to
compare the energies of one and two layers in the box, red and
green curves in Figure \ref{Figure-10B-Energy}, correspondingly.
The energies are quite similar at low densities of the layer and
diverge significantly at high densities. This is consequence of
the compression of the two layers due to lack of space in the box
for two layers of equilibrium size. The energy per lipid
corresponding to the maximal compression of two layers in the box
coincides with the energy of a compressed layer. This fact
reflects the consistency of our picture. On the other hand, high
compression of a single layer, $A$ below $50$\AA $^{2}$, results
in gradual splitting of the layer in two with smooth rearrangement
of the lipid molecules. Some phospholipids flip and orient their
\textquotedblleft heads\textquotedblright\ in the center of the
membrane. Thus, a single bilayer smoothly become broader and
change its appearance to remind something between a single bilayer
with a layer of heads\ in the center and two tightly compressed
bilayers. This process is reflected in the flat part of the energy
per lipid curve for $A$ below $50$\AA $^{2}$, where the
distinction between one layer and two layers in the box can hardly
be done.

The energy curve for the 3-beads model is presented in Figure \ref%
{Figure-3B-Energy}. This is the simplest phospholipid model with
only 3 beads, thus the conformational space is very limited and
the results obtained with the sampling of 1 million of
conformations represented here are close to these obtained with
small sampling about 100-300 thousands. The difference between the
energies of one and two layers in the box, noted for 10-beads
model, is even more pronounced.

The mean field "snapshots" of the most probable conformations of the layer
in equilibrium and the corresponding density profiles for three models are
presented in Figure \ref{Figure-Snaps}. It shows how the coarse-graining and
the decreasing of the model details influence the layer appearance. However,
despite obvious differences in details caused by the differences in the
molecule structures, the profiles looks quite similar, especially for the
most coarse-grained 10- and 3-beads models. In case of the 44-units model,
there is a small fraction of the hydrophobic T units at the surface,
reflecting the structure of the phospholipid with some hydrophobic T groups
in the head. Furthermore, different aspect ratio between T and H units in
the models ($4:1$ in 10-beads and $2:1$ in the 3-beads model) lead to the
corresponding difference in the relative density of T and H units. This
correspondence is not valid for the 44-beads model, where the beads do
overlap and some parts of the molecule are hidden from the solvent. The
ratio between the areas under the T and H curves is $4.7:1$ which is
different from the ratio between numbers of T and H beads in the molecule ($%
3.4:1$).

Finally few words should be said about the speed performance of
three models. We run six series of calculations with different
conformational sampling of 1-3 millions of conformations. The box
is divided in $\sim 100$ parallel layers and the number of lipids
in the box is varying from $300$ to $700$ with the step $10$. The
program runs in parallel on 32-cores AMD machine with 128Gb of
RAM. The most time consuming process is the solution of equations
while the generation of the sampling at the beginning of
calculation is relatively fast and almost does not affect the
speed of calculation. In turn, the solution of the equations
depends on the number of conformations in the sampling, the number
of cells and the number of attempts to solve equations with
different initial conditions. Since three models differ in the
number of beads, implying different conformational space, the same
accuracy of calculations in different models is achieved with
different number of conformations in the sampling. The 44-beads
model require several millions of conformations and 48 hours of
calculations, the 10-beads model require one million of
conformations and 17 hours, while the 3-beads model can get
reliable results with a hundred thousands of conformations in few
hours. Thus, the simplest 3-beads model is the fastest model which
can reproduce the essential equilibrium properties of the
phospholipid bilayer.

\section{Conclusions \label{secsconclusions}}

In this work we report a computational tool providing relatively fast and
stable solution of the equations of the SCMF theory in different geometries
and different molecule structures. The SCMF theory is applied for the first
time to model the DMPC phospholipid membranes with realistic thermodynamic
properties. We show that all three coarse-grained models of phospholipids
with the present parameters
can adequately describe the equilibrium properties and the molecular
structure of DMPC phospholipid bilayers in fluid phase. Among the essential
properties of the phospholipid bilayers that correspond to the experimental
data are the thickness of the membrane, the positions of different groups of
the lipid in the bilayer, the compressibility of the membrane and the
equilibrium interfacial area per lipid. The most detailed 44-beads model
can reproduce the structure of the phospholipid layer with maximum details. The
resulting density profiles of the groups of the phospholipid molecule can be
used for the structural modeling in experimental techniques that require the
molecular structure of the layer.

However, if one is interested in the average properties of the
lipid bilayer, the simplest 3-beads model can be used to model the
thermodynamics and the essential properties of self-assembled
phospholipid bilayers. This model is faster and computationally
more attractive than 44- and 10-beads models. Furthermore, we
suggest that the 3-beads model with the parameters found in this
work is used for simulation of more complex bilayer structures.

\section*{Acknowledgments}

The authors are grateful to Josep Bonet and Allan Mackie for
introducing into the method SCMF theory, Nigel Slater for his
hospitality during the visit of the University of Cambridge and
stimulating discussions. The authors gratefully acknowledges
financial help from Spanish Ministry of education MICINN via
project CTQ2008-06469/PPQ and the UK Royal Society for the
International Joint Project.


\appendix

\section{Computational Details \label{seccompdet}}

Here we discuss the implementation details of the algorithm underlying the
computational realization of the SCMF theory. The off-lattice realization of
the program allows for conformations of the molecules in a real space. The
algorithm consist of three independent parts: the generation of the sampling
of conformations of molecules of any structure, the discretization of space
according to the geometry of the problem and the solution of the SCMF
equations (\ref{MainEquations}). The independence of the parts insures the
universality of the method which can be applied for molecules of any
structure and their self-assembly in objects of any geometry.

\subsection{Conformational sampling generation\label{secgeneration}}

First step is the generation of a representative set of conformations of a
single molecule. This can be done with different techniques, including
bead-to-bead chain growth or MC simulations. The conformations of a single
molecule are generated once before solving the equations. They are stored in
the RAM memory and not changed during the solution of equations, while the
probabilities of each conformation are recalculated with the fields. This
static memory allows for highly efficient parallelization of the program
within the shared memory OpenMP platform, since the processors communicate
very little between each other. The generation of conformations in the
sampling is realized as follows:

\begin{enumerate}
\item Generation of a new conformation $\Gamma _{\alpha }$ of the molecule
of type $\alpha $ with random position of the first bead inside the box. We
set initial values of the conformation statistical weight $w_{\alpha
}(\Gamma _{\alpha })=1$ and the intramolecular energy $H_{\alpha
}^{intra}(\Gamma _{\alpha })=0$.

\item Iterative addition of new beads to the growing molecule and joining
them one by one subject to self-avoidance condition. Depending on the final
objective, one can use:

\begin{enumerate}
\item \label{JoinFixPos}Self-avoiding growth with the MC equilibration. The
chain is generated as a self-avoiding off-lattice walk and the energy of
interaction of the newly joined beads with the rest of the chain is
accumulated in $H_{\alpha }^{intra}(\Gamma _{\alpha })$. The resulting chain
is then equilibrated with MC crankshaft moves and reptations until the full
equilibration of the configuration. In this case all conformations have the
same statistical weight and $w_{\alpha }(\Gamma _{\alpha })=1$.

\item \label{JoinFixBond} Rosenbluth chain growth \cite{Rosenbluth}. A new
bead is placed at a fixed distance from the previous bead either at random
angle (freely joined chain) or with additional restrictions on the angles
between the beads (chain with angle restrictions). Rosenbluth algorithm
generates conformations with biased probability distribution. This bias is
then calculated in the conformation weight and removed when the
probabilities of each conformation are calculated. In order to calculate the
Rosenbluth weight, one should try $N_{trial}$ times to place the new bead at
a fixed distance from the previous bead and calculate the number $%
N_{allowed} $ of successful positions, allowed by self-avoidance
condition. If $N_{allowed}>0$, a new position is accepted with the
weight $1/N_{allowed} $, the energy of interaction is accumulated
in $H_{\alpha }^{intra}(\Gamma _{\alpha })$ and the weight of the
conformation $w_{\alpha }(\Gamma _{\alpha })$ is multiplied by the
factor $1/N_{allowed}$. If there is no possibility to place the
bead, $N_{allowed}=0$, the generation restarts from the beginning.
The Rosenbluth algorithm is simple and efficient enough for
generation of phospholipid molecules.
\end{enumerate}

\end{enumerate}

\subsection{Discretization of space}

Since the solution of equations of the SCMF theory in the integral form (\ref%
{MainEquations}), (\ref{FractionOfSolvent}) is not practical, the equations
are simplified by the replacement of the integrals with the sums. The space
is discretized according to the geometry of the system and the symmetry
considerations. The integrals over the spatial coordinates, $d\pmb r$, are
replaced by the sums over auxiliary cells $i$ with the same values of the
mean fields: $\left\langle c_{\beta }^{a}(\pmb r)\right\rangle =c_{\beta
i}^{a}$ and $\lambda (\pmb r)=\lambda _{i}$. The discretization leads to the
following approximate relations
\begin{equation}
\begin{split}
\int u_{\alpha }^{a}(\Gamma _{\alpha },\pmb r)\left\langle c_{\beta }^{a}(%
\pmb r)\right\rangle d\pmb r& \approx \sum_{i}c_{\beta
i}^{a}\int_{i}u_{\alpha }^{a}(\Gamma _{\alpha },\pmb r)d\pmb r \\
\int \lambda (\pmb r)\phi _{\alpha }(\Gamma _{\alpha },\pmb r)d\pmb r&
\approx \sum_{i}\lambda _{i}\int_{i}\phi _{\alpha }(\Gamma _{\alpha },\pmb %
r)d\pmb r
\end{split}
\label{CellsDivision}
\end{equation}%
where $\sum_{i}$ denotes the sum over all auxiliary cells, while $\int_{i}$
is the integral over the $i$-th cell. Because of the structure of the
equations, the integral over the single $i$-th cell can be evaluated once in
the beginning of calculations, while the sums (\ref{CellsDivision}) are
calculated during the solution of the equations. Although the number and the
geometry of the cells can be arbitrary, their choice will restrict the
geometry and the resolution of the resulting solutions. Thus, it is
important to choose them accurately for a particular problem. In case of
flat bilayers we choose a planar geometry with the cells parallel to the
layer, thus we can only obtain the solutions that can lie in the layer.

The integrals over the conformational space of single molecules are replaced
by the sums over the finite samplings generated in the first step
\begin{eqnarray}
\phi _{\alpha i} &=&\sum_{\Gamma _{\alpha }}\phi _{\alpha i}(\Gamma _{\alpha
})\rho _{\alpha }(\Gamma _{\alpha }),  \label{MainEquationsDiscrete} \\
c_{\alpha i}^{a} &=&\sum_{\Gamma _{\alpha }}c_{\alpha i}^{a}(\Gamma _{\alpha
})\rho _{\alpha }(\Gamma _{\alpha })\quad  \notag
\end{eqnarray}%
where the mean field volume fraction $\phi _{\alpha i}$ and the mean field
concentration $c_{\alpha i}^{a}$ of $a$-beads of the molecule of type $%
\alpha $ in the cell $i$ are related to the corresponding volume fraction $%
\phi _{\alpha i}(\Gamma _{\alpha })$ and the concentration $c_{\alpha
i}^{a}(\Gamma _{\alpha })$ of the conformation $\Gamma _{\alpha }$.

The integral expression (\ref{ProbabilityExpression}) for the $\rho _{\alpha
}(\Gamma _{\alpha })$ yields in form
\begin{equation}
\begin{split}
\rho _{\alpha }(\Gamma _{\alpha })& =\frac{1}{Z_{\alpha }w_{\alpha
}(\Gamma _{\alpha })}\exp \left( -H_{\alpha }^{intra}(\Gamma
_{\alpha })- \right.\\
&\sum_{\beta
=1}^{M}(N_{\beta }-\delta _{\alpha \beta })\times  \\
& \left. \times \sum_{ai}\epsilon _{\alpha i}^{a}(\Gamma _{\alpha })c_{\beta
i}^{a}+\sum_{i}\lambda _{i}v_{\alpha i}(\Gamma _{\alpha })\right)
\end{split}
\label{ProbabilityExpressionDiscrete}
\end{equation}%
where $w_{\alpha }(\Gamma _{\alpha })$ is the Rosenbluth weight calculated
during the generation of the chains, and the constant $Z_{\alpha }$ insures
the normalization of the probability $\sum_{\Gamma _{\alpha }}\rho _{\alpha
}(\Gamma _{\alpha })=1$.

The expression for probabilities (\ref{ProbabilityExpressionDiscrete}) is
accompanied with the following relations
\begin{equation}
\begin{split}
v_{s}\lambda _{i}& \approx \ln \phi _{si}+\sum_{\beta
=1}^{M}(N_{\beta }-\delta _{s\beta })\sum_{a}\epsilon
_{si}^{a}c_{\beta i}^{a},\\
&\quad \phi _{si}=\phi _{0}-\sum_{\alpha =1,\alpha \neq
s}^{M}N_{\alpha }\phi _{\alpha i}
\\
\epsilon _{\alpha i}^{a}(\Gamma _{\alpha })& =\int_{i}u_{\alpha
}^{a}(\Gamma _{\alpha },\pmb r)d\pmb r,\\
&\quad v_{\alpha i}(\Gamma _{\alpha })=V_{i}\phi _{\alpha
i}(\Gamma _{\alpha })=\int_{i}\phi _{\alpha }(\Gamma _{\alpha
},\pmb r)d\pmb r
\end{split}
\label{AdditionalExpressionsDiscrete1}
\end{equation}%
The quantities $v_{\alpha i}(\Gamma _{\alpha })$, $\epsilon _{\alpha
i}^{a}(\Gamma _{\alpha })$ and $c_{\alpha i}^{a}(\Gamma _{\alpha })$ are
calculated using the MC integration technique. The volumes $v_{\alpha
i}(\Gamma _{\alpha })$ are calculated as follows: one throws randomly $%
N_{trials}$ points inside the volume occupied by the bead and determine the
cell where each of these points hits. Each point contributes to the volume
in the cell with the value $v_{a}/\left( N_{trials}N_{int}\right) $, where $%
N_{int}$ is the number of the beads hit by the point and $v_{a}$ is the
volume of the bead. The volume fractions $\phi _{\alpha i}(\Gamma _{\alpha })
$ are related to the volumes $v_{\alpha i}(\Gamma _{\alpha })$ as $\phi
_{\alpha i}(\Gamma _{\alpha })=v_{\alpha i}(\Gamma _{\alpha })/V_{i}$, where
$V_{i}$ is the volume of the $i$-th auxiliary cell. The concentrations $%
c_{\alpha i}^{a}(\Gamma _{\alpha })$ are calculated similarly, except that
each random point contributes to $c_{\alpha i}^{a}(\Gamma _{\alpha })$ with
the value $1/N_{trials}$. The interaction energies $\epsilon _{\alpha
i}^{a}(\Gamma _{\alpha })$ are calculated with the help of the $N_{trials}$
points thrown for every $a$ at a distance $r$ from the center of the bead $b$
such that $r_{min}<r<r_{max}$, where $r_{min}$ is the minimum distance at
which the bead of type $a$ can approach to the $b$ bead, and $r_{max}$ is
the maximum distance at which the interaction potential acting between $a$
and $b$ beads is non-zero. If the bead of type $a$ in the position of the
trial point does not have any intersection with the beads of the chain, its
contribution to the $\epsilon _{\alpha i}^{a}(\Gamma _{\alpha })$ is $\frac{4%
}{3}\pi \left( r_{max}^{3}-r_{min}^{3}\right) U^{ab}(r)/N_{trials}$, where $i
$ is the number of the auxiliary cell where the trial point hits, and $%
U^{ab}(r)$ is the pair potential acting between the beads of types $a$ and $b
$. Since the solvent is implicit in our model, the quantities $\epsilon
_{si}^{a}$ describing the interaction between the solvent and the bead of
type $a$ have to be treated separately. We evaluate them by a simple
approximate expression
\begin{equation}
\epsilon _{si}^{a}\approx \frac{4}{3}\pi
(r_{sa}^{3}-(r_{s}+r_{a})^{3})\varepsilon _{sa}
\label{AdditionalExpressionsDiscrete2}
\end{equation}%
where we assume the square well potentials between the solvent molecules $s$
and the beads $a$ with energy $\varepsilon _{sa}$ at the distance closer
than $r_{sa}$. Here the radius of solvent is $r_{s}$ and the radius of the $a
$ bead is $r_{a}$.

It is noteworthy that the incorporation of the Rosenbluth weight $w_{\alpha
}(\Gamma _{\alpha })$ into the expression for probabilities (\ref%
{ProbabilityExpressionDiscrete}) may lead to confusion. Although most of the
averages can be calculated with the expressions similar to (\ref%
{MainEquationsDiscrete}), the averages of expressions containing explicitly
the probability $\rho _{\alpha }$ should be treated separately. For example,
the entropy term in the free energy (\ref{FreeEnergyFull}) leads to the
additional term when we replace the integral with the sum
\begin{eqnarray}
&N_{\alpha }\left\langle \ln {\frac{\rho _{\alpha }N_{\alpha
}}{e}\Lambda _{\alpha }}\right\rangle =N_{\alpha }\sum_{\Gamma
_{\alpha }}\rho _{\alpha }(\Gamma _{\alpha })\ln {\frac{\rho
_{\alpha }(\Gamma _{\alpha })N_{\alpha }}{e}\Lambda _{\alpha }}+\nonumber\\
&N_{\alpha }\sum_{\Gamma _{\alpha }}\rho _{\alpha }(\Gamma
_{\alpha })\ln {w_{\alpha }(\Gamma _{\alpha })\varkappa _{\alpha
}}
\end{eqnarray}%
where $\varkappa _{\alpha }$ is the number of conformations in the
sampling of the molecule of type $\alpha $.

The expressions (\ref{MainEquationsDiscrete}), (\ref%
{ProbabilityExpressionDiscrete}) and (\ref{AdditionalExpressionsDiscrete1})
form a closed set of algebraical equations with respect to variables $\phi
_{\alpha i}$ and $c_{\alpha i}^{a}$ which can be solved numerically. We must
admit however, that the solution of these non-linear equations is not a
straightforward task.

\subsection{Numerical implemetation\label{secsolution}}

In this work we have implemented the SCMF theory in C++. The program is
based on Newton-Rapson technique which provides the efficient and stable
solution of the SCMF equations. It has a modular structure which makes it
very flexible for modification, further development and extension for new
applications.

The first module is the sampling generator, which generates the molecules
conformations and calculates the volumes and interactions according to the
Appendix \ref{secgeneration}. The code for generation of molecules with
different structures can be added to this module or changed independently
from other parts of the program.

The second module specify the geometry of the problem: 1D spherical, 1D
planar, 2D planar, 2D cylindrical, or 3D cubic. It contains the description
of the geometry and divide the simulation box into set of auxiliary cells.
The modification of this part of code provides an easy way to implement
different types of the system geometry. It is worth to note here one
technical improvement that significantly increase the performance of the
calculations. Since the molecules have small size compared to the size of
the box, their contribution to most of the auxiliary cells is zero, thus
matrices $\epsilon _{\alpha i}^{a}(\Gamma _{\alpha })$, $c_{\alpha
i}^{a}(\Gamma _{\alpha })$ and $v_{\alpha i}(\Gamma _{\alpha })$ have many
zero values. In this second module of the program the simple compression of
these data is realized via elimination of zeros from these matrices. This
helps to save computer memory and, on the next step, speed up the
calculations by elimination of the sums over the cells containing zeros.

The core of the third module is the modified Newton-Rapson solver for the
solution of SCMF equations. It involves the iterative solver which
calculates on each step the matrix of analytical derivatives of equations %
\ref{MainEquationsDiscrete} with respect to unknown variables along with the
analytical expressions including the sums over the conformational samplings.
The next approximation of the solution is determined by the inversion of
this matrix until the required accuracy is reached. To avoid occasional
instability of the Newton-Rapson scheme, the algorithm artificially
restrains too big steps which may drop the next approximation outside a
reasonable range. The algorithm is able to find several solutions
simultaneously by repeating the calculations several times with different
initial conditions.

\pagebreak
\bibliographystyle{rsc}
\bibliography{final-version}

\end{document}